\documentclass[prd,preprint,superscriptaddress,amsmath,amssymb,nofootinbib]{revtex4}

\usepackage{bibentry}
\usepackage{graphicx,color,slashed}

\def\be{\begin{eqnarray}}
\def\ed{\end{eqnarray}}

\begin{document}


\title{\bf \large Optimizing $pp\to A\to Z^{*}h\to \ell^+\ell^- b\bar b$ Searches at the LHC\\ in the 2HDM Type-I with Inverted Hierarchy}

\author{A.G. Akeroyd}
\email{a.g.akeroyd@soton.ac.uk}
\affiliation{School of Physics and Astronomy, University of Southampton,
Highfield, Southampton SO17 1BJ, United Kingdom}

\author{S. Alanazi}
\email{swa1a19@soton.ac.uk; SWAlanazi@imamu.edu.sa}
\affiliation{School of Physics and Astronomy, University of Southampton,
Highfield, Southampton SO17 1BJ, United Kingdom}
\affiliation{Physics Department, Imam Mohammad Ibn Saud Islamic University (IMISU), P.O. Box 90950, Riyadh, 11623, Saudi Arabia}

\author{S. Moretti}
\email{stefano.moretti@cern.ch}
\affiliation{School of Physics and Astronomy, University of Southampton,
Highfield, Southampton SO17 1BJ, United Kingdom}
\affiliation{Department of Physics and Astronomy, Uppsala University, Box 516, 751 20 Uppsala, Sweden}


\begin{abstract}
{\noindent\footnotesize
In this study, we investigate the Large Hadron Collider (LHC) search for the signal process $pp\to A \to Z^{*}h \to \ell^{+} \ell^{-} b \bar{b}$ ($\ell=e,\mu$) within the framework of the 2-Higgs-Doublet Model (2HDM) Type-I, considering an Inverted Hierarchy (IH) scenario wherein the Standard Model (SM)-like Higgs boson $H$ is heavier than $h$ (i.e., $m_h < m_H=125$ GeV). We reproduce the dominant background distributions from a CMS analysis of $pp\to H \to Z A \to  \ell^{+} \ell^{-} b \bar{b}$ (for a on-shell $Z$), which we do for validation purposes, so that we can explore different invariant mass selection criteria to enhance the signal significance of our target process. Specifically, we compare the CMS baseline cuts ($70$ GeV $ < m_{\ell^+\ell^-} < 110$ GeV, no $m_{b\bar b}$ restriction) with the alternative selections $20~{\rm GeV}~< m_{\ell^+\ell^-} < 50$ GeV with $m_{b\bar b} < 100$ GeV. 
The latter cuts are enforced to account for the off-shellness of the $Z^*$ boson 
and the low mass Higgs state 
in our case. We show that these modifications reduce drastically  the dominant Drell-Yan (DY) and top-(anti)quark pair backgrounds, leading to a significant excess  from the analysis of the reconstructed $m_{\ell^+\ell^-b\bar b}$ invariant mass in the illustrative ranges 145 GeV $<m_{A}<$ 150 GeV and 68 GeV $<m_{h}<$ 75 GeV. Our results demonstrate that such  cuts enable effective signal discrimination, suggesting an optimized strategy in future searches for an established topology but in a new kinematical regime, which is of particular relevance to the 2HDM Type-I with IH in its mass spectrum.}
\end{abstract}

\maketitle

\section{Introduction}
\label{Sec:Intro}
\noindent 
The discovery of a Higgs boson in 2012 at the LHC marked a significant milestone in particle physics, yet the possibility of an extended Higgs sector remains an open question. One well-motivated extension of the SM is the 2HDM, which introduces several (pseudo)scalar Higgs states: two charged ones $H^{\pm}$, one CP-odd one $A$ and two CP-even ones $h$ and $H$ (with, conventionally, $m_{h}<m_{H}$). Because $h$ and $H$ have  zero electric charge and exhibit the same parity, this opens the door for a scenario where one of these two states matches the discovered one. To control Flavor-Changing Neutral Currents (FCNCs), a discrete 
$\mathbb{Z}_2$ symmetry is customarily applied to the 2HDM. This leads to four so-called Yukawa types of it: Type-I, -II, -Flipped and -Lepton Specific, depending on how this symmetry is assigned to the fermions and Higgs doublets \cite{Branco:2011iw}.

In this theoretical context, we explore the IH scenario, wherein the heavier CP-even neutral Higgs boson $H$ corresponds to the discovered Higgs state, in the 2HDM Type-I (i.e., $m_H =125$ GeV). Under this condition, the production and decay mode $pp\to A \to Z^{(*)}h$ provides an interesting channel for probing such a new physics, where the weak gauge boson can be either on- ($Z$) or off-shell ($Z^*$). A possible final state emerging from this process is $\ell^+\ell^- b\bar b$, due to further $Z^{(*)}$ and $h$ decays, respectively, which has been studied in several phenomenological and experimental papers. However, in the spirit of Ref.~\cite{Semlali:2020cxl}, and based on Ref.~\cite{CMS:2019ogx}, {which studied the processes $pp\to H \to ZA $ and $pp\to A  \to ZH $ in the $\ell^+\ell^- b\bar b$ final state at $\sqrt{s}=13$ TeV with 35.9 fb$^{-1}$ of CMS data, we replicated the dominant background contributions (DY and $t\bar t$) from that analysis. This places us in the position of addressing the possibility of extracting our target signal in the complementary mode $pp\to A\to Z^{(*)}h$, i.e., with $m_A<215$ GeV.} 

Therefore, in this short letter, based on previous results of ours on the signal \cite{Akeroyd:2023kek,Akeroyd:2024dzq,Akeroyd:2024tbp}\footnote{We note that a similar study to ours (chiefly, exploiting the off-shellness of the $Z^*$ boson) was already carried out in Ref.~\cite{Accomando:2020vbo}, albeit for a 2HDM Type-II wherein $h$ was the SM-like Higgs boson.}, 
we investigate the process $pp\to A \to Z^*h \to \ell^+\ell^-b\bar b$ and explore optimized selection criteria that enhance its significance. We consider the two aforementioned main backgrounds, DY ($q\bar q\to \gamma^*,Z\to \ell^+\ell^-$) and top-(anti)quark pair production ($gg,q\bar q\to t\bar{t}$ with two leptonic decays of the emerging $W^\pm$ pair). By comparing different invariant mass selection criteria, we demonstrate that specific mass cuts can significantly suppress such backgrounds while retaining the majority of signal events, ultimately improving the LHC sensitivity to the $pp\to A \to Z^*h$ channel in the 2HDM Type-I scenario with IH, crucially, for the case of an off-shell $Z^*$. 

{Experimental searches for new Higgs bosons in the $\ell^+\ell^- b\bar b$ final state have mainly targeted the process $pp\to A \to Zh_{\rm SM}$, under the assumption that the lighter CP-even state $h$ of the 2HDM is the observed 125 GeV Higgs boson $h_{\rm SM}$ (standard mass  hierarchy) and that the $Z$ is produced on-shell. Specifically, the CMS search at 13 TeV with 35.9 fb$^{-1}$ of Ref.~\cite{CMS:2019qcx} probed the mass range $m_{A}=225$–-$1000$ GeV while the corresponding ATLAS analyses at 8 TeV  with 20.3~fb$^{-1}$ \cite{ATLAS:2015} and 13 TeV with 139 fb$^{-1}$ \cite{ATLAS:2021} explored the range $m_{A}=220$–-$800$ GeV. 
By contrast, Ref.~\cite{CMS:2019ogx} studied the related processes $pp\to H \to ZA$ and $pp\to A \to ZH$ in the same final state, using the 2016 CMS dataset at 13 TeV with 35.9 fb$^{-1}$. In this case the two additional neutral Higgs bosons (i.e., non-SM-like) of the 2HDM were searched for in the ranges $120$ GeV $<m_{H}<1000$ GeV and $30$ GeV $<m_{A}<1000$ GeV, respectively. 
The last analysis is particularly relevant to our work because it also allows for the IH interpretation when the heavier CP-even state of the 2HDM  plays the role of the discovered 125 GeV Higgs. 
However, the mass range explored in Ref.~\cite{CMS:2019ogx} implies that the $Z$ boson is always on-shell (i.e., $m_{A}>m_Z+m_H$  in the $pp\to A\to ZH$ channel).  
}

The plan of the paper is as follows. In the next section, we describe our methodology. We then move on to present our results before concluding in the last section.

\section{Methodology}
\label{72}
This section provides an overview of the two largest backgrounds to the signal process $pp\to A \to Z^*h\to \ell^+\ell^-b \bar{b} $,  
which are DY and top-(anti)quark pair production. The search in  \cite{CMS:2019ogx} is carried out 
 with LHC Run 2 $pp$ collision data corresponding to an integrated luminosity of $35.9$ fb$^{-1}$, which was collected by the CMS experiment.  We computed the signal $pp\to A \to Z^*h \to \ell^+\ell^-b\bar{b}$  for two Benchmark Points (BPs) within the framework of the 2HDM Type-I, considering the IH scenario with  $m_{H} = 125$ GeV,  $m_{h}< 125$ GeV and $m_A$ chosen in the range $145$ GeV 
 $< m_{A} < 150$ GeV, by setting  $m_{H^\pm} = m_{A}$ in order to comply with measurements of Electro-Weak Precision Observables (EWPOs). We finally choose $m_{A}$ and $m_{h}$  so that the $Z^*$ boson remains off-shell, i.e., $m_A-m_h<m_Z$.
 
Regarding the (cosine  of  the) mixing angle $\cos(\beta - \alpha)$, since we are working within the IH scenario, this is required to be very close to 1 \cite{Akeroyd:2023kek,Akeroyd:2024dzq,Akeroyd:2024tbp}. In our study, for simplicity, we have fixed $\cos(\beta - \alpha)=1$ exactly.
Furthermore, to ensure both theoretical consistency and compatibility with experimental constraints, we have considered values of the ratio of the 2HDM Vacuum Expectation Values (VEVs) in the range $2.5 < \tan\beta < 10$. 

By doing so, we have fixed all the free parameters of the CP conserving 2HDM that we assume here, wherein only soft breaking of the discussed $\mathbb{Z}_2$ symmetry is allowed, as opposed to a hard breaking of it (i.e., $m_{12}\ne 0$ and $\lambda_{6,7}=0$ in the 2HDM Lagrangian \cite{Branco:2011iw}).

\subsection{The DY and  $t\bar{t}$ Backgrounds}
\begin{figure}
    \centering
    \includegraphics[width=0.9\linewidth]{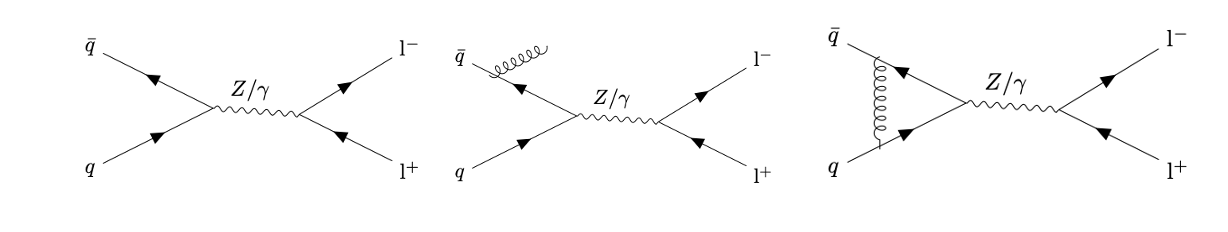}
    \caption{Representative Feynman diagrams for DY production at LO and in presence of QCD radiation.}
    \label{DY}
\end{figure}

\begin{figure}
    \centering
    \includegraphics[width=0.9\linewidth]{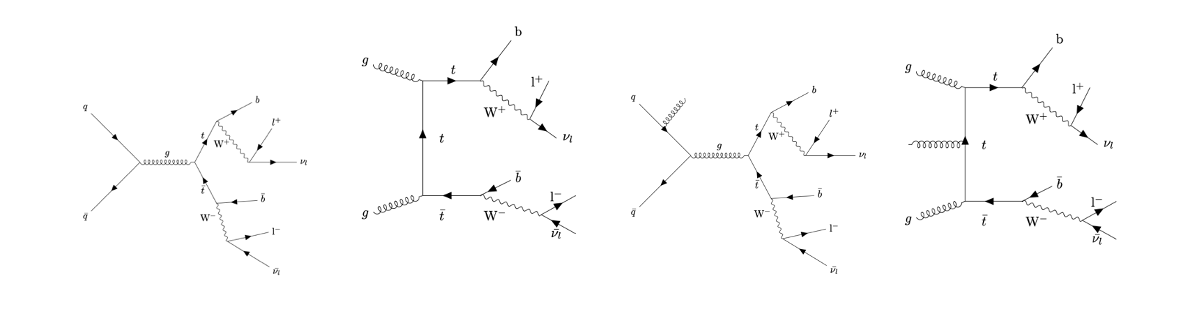}
    \caption{Representative Feynman diagrams for $t\bar{t}$ production at LO and in presence of QCD  radiation.}
    \label{TT}
\end{figure}

The DY process and top-(anti)quark pair production and decay are critical backgrounds in numerous analyses at the LHC, particularly in searches for new physics. Sample Feynman diagrams are shown in
Figs.~\ref{DY}--\ref{TT}, respectively. The first process occurs when a light quark and  antiquark from the protons annihilate, producing a virtual photon or $Z$ boson, which then decay into a pair of leptons ($\ell^+\ell^-$) and initial state radiation fakes the $b\bar b$ system.  The second one is primarily induced by gluon-gluon fusion (with only a subleading component from $q\bar q$ annihilation), where a top quark $t$ and its antiparticle $\bar{t}$ are created together and then decay via $t\bar{t}\to W^+W^-b\bar{b}\to \ell^+\ell^-b\bar{b}$ plus missing transverse energy/momentum.

Both processes can be generated at different orders in perturbation theory: i.e., at Leading Order (LO) or in presence of higher order corrections where Quantum Chromo-Dynamics (QCD) and/or EW radiative effects are introduced. The ATLAS and CMS collaborations have carried out comprehensive studies of these two processes across two center-of-mass energies of the LHC, i.e., using the 8 and 13 TeV datasets, for  varying values of integrated luminosity (see, e.g.,  \cite{ATLAS:2019dy}, \cite{ATLAS:2025dy} and \cite{CMS:2023ttbar}).
  
  As already mentioned, we have finally used Ref.~\cite{CMS:2019ogx} to calibrate our Monte Carlo (MC) simulation of the two dominant background processes. 


\subsection{Parameter Space Sampling, Event Generation and Mass Reconstruction}

We selected two BPs  for simulating our signal $pp\to A\to Z h^{*}\to \ell^+\ell^-b \bar{b} $, hereafter, labelled as BP1 and BP2. They have been tested and found to be consistent with current theoretical and experimental constraints, by using HiggsTools \cite{Bahl_2023}. Flavor constraints were checked using SuperISO \cite{Mahmoudi:2008tp}. For BP1, we use $m_{A}=m_{H^{\pm }}= 150 \, {\rm GeV}, \, m_{H}= 125 \, {\rm GeV}, \, m_{h}= 68 \, {\rm GeV},\, \tan\beta = 3.9$ whereas for BP2 we use $m_{A}=m_{H^{\pm }}= 145 \, {\rm GeV}, \, m_{H}= 125 \, {\rm GeV}, \, m_{h}= 75 \, {\rm GeV},\, \tan\beta = 4$. The mass difference relevant in each case is $m_{A}-m_{h}=82$ GeV in BP1 and 70 GeV in BP2, thus precluding an on-shell $Z$ boson. 

We followed the same methodology employed by the CMS collaboration \cite{CMS:2019ogx} to generate the DY and  $t\bar{t}$ backgrounds. That is, the
corresponding events were generated at Next-to-LO (NLO) using  MadGraph5aMC@NLO \cite{Alwall2014} with the FxFx merging scheme \cite{Frederix_2012}. The latter enables consistent merging of multi-jet NLO samples and avoids double counting between matrix elements and parton shower emissions. MadSpin was then used to decay the $W^\pm$ and $Z^{(*)}$ bosons to di-leptons. The events were then passed to PYTHIA \cite{Sjostrand:2014zea} for parton showering and hadronization, employing the NNPDF 3.0 Parton Distribution Function (PDF) set \cite{PDF} with the (dynamical) factorization/renormalization scale set by MadGraph5aMC@NLO. Then for all processes we applied detector simulation via Delphes \cite{deFavereau:2013fsa} with a standard CMS card \cite{Mertens:2015kba}. We have then used the same tools and settings for our signal emulation, including the jet merging feature.

Selection cuts were applied on the transverse momentum ($p_{T})$, rapidity ($\eta $) and separation ($\Delta R$) of (some of) the final-state particles as well as the di-lepton ($m_{\ell^+\ell^-}$) and di-jet ($m_{b\bar b}$) invariant masses. 
We used MadAnalysis5 \cite{Conte2013} to manipulate the final state objects, two $b$-jets and electron/muon pairs. Events with same flavor leptons $(e^{\pm}e^{\mp},\mu^{\pm}\mu^{\mp})$ are allowed in the signal definition whereas  different flavour leptons $(e^{\pm}\mu^{\mp})$  are not.  Moreover, for jets, we have applied jet reconstruction by using the anti-$k_{T}$ algorithm with a cone parameter of 0.4, as implemented in the FastJet package \cite{Cacciari_2012}.

\section{Results}
\label{73}

The cuts used initially, based on  Ref.~\cite{CMS:2019ogx}, were as follows:
\begin{itemize}
    \item $p_{T}^{e^{\pm }}> 25$ GeV (leading), 15 GeV (subleading) and $\left| \eta (e^{\pm })\right|<2.5$;
    \item $p_{T}^{\mu^{\pm }}> 20$ GeV (leading), 10 GeV 
    (subleading), and $\left| \eta (\mu^{\pm })\right|<2.4$;
    \item $p_{T}^{e^{\pm }\mu^{\mp }}> 25$ GeV (leading), 15 GeV (subleading $e^{\pm }$), 10 GeV (subleading $\mu^{\pm }$);
    \item $p_{T}^{j}>20$ GeV and $\left| \eta (j)\right|<2.4$;
    \item $\Delta R_{jj}>0.4$ and $\Delta R_{\ell^+\ell^-}>0.3$;
    \item $70$ GeV $< m_{\ell^+\ell^-}< 110$ GeV.
    
\end{itemize}

\begin{figure}
    \centering
    \includegraphics[width=0.8\linewidth]{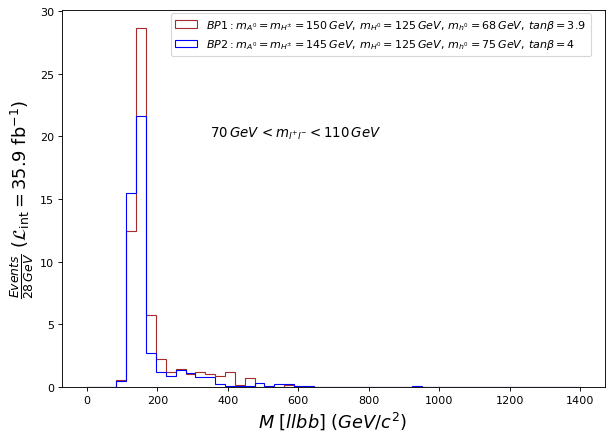}
    \includegraphics[width=0.8\linewidth]{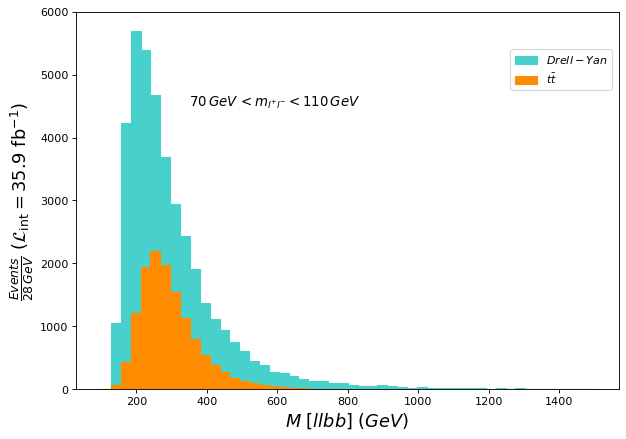}
    \caption{Signal and the background processes as a function of $m_{\ell^+\ell^-b\bar b}$ for $70$ GeV $< m_{\ell^+\ell^-} < 110$ GeV.}
    \label{SSBG70}
\end{figure}

First, we investigated the significance of the signal for the two BPs with the CMS standard cuts as listed above. As intimated, we have included only the two dominant backgrounds, which are the DY and $t\bar{t}$ processes. In
Fig.~\ref{SSBG70} the mass distribution $m_{\ell^+\ell^-b\bar b}$ is analysed using a mass binning width of 28 GeV, as in \cite{CMS:2019ogx}. This allows us to search for potential excesses of events for each signal BP and to assess the shape of the signal in comparison to the DY and $t\bar{t}$ backgrounds. A similar plot has been shown in the CMS search in \cite{CMS:2019ogx}, but with different signal BPs. Our results for the background processes 
(lower plot in  Fig.~\ref{SSBG70}) agree very well with the corresponding plot in \cite{CMS:2019ogx}, with the peak at around 200 GeV for DY and at around 250 GeV for $t\bar{t}$. For the two signal BPs one sees that the peak in 
$m_{\ell^+\ell^-b\bar b}$ is at
the mass of $m_{A}$, as expected, but the number of signal events is much less than that of the background. Hence there would be no possibility of observing a statistical significant signal  in our case, where the primary decay is $A\to Z^*h$. We in fact note that the cut  $70$ GeV $< m_{\ell^+\ell^-}< 110$ GeV was employed by 
CMS in \cite{CMS:2019ogx} in order to
keep most of the signal events in the decay $H\to ZA$ with a real $Z$. However, this cut removed a large fraction of the signal events for our BPs with an off-shell $Z^*$, as the $m_{\ell^+\ell^-}$ 
distribution from $Z^*$ would not be expected to concentrate in the region $70$ GeV $< m_{\ell^+\ell^-}< 110$ GeV (for our BPs).

\begin{figure}
    \centering
    \includegraphics[width=0.8\linewidth]{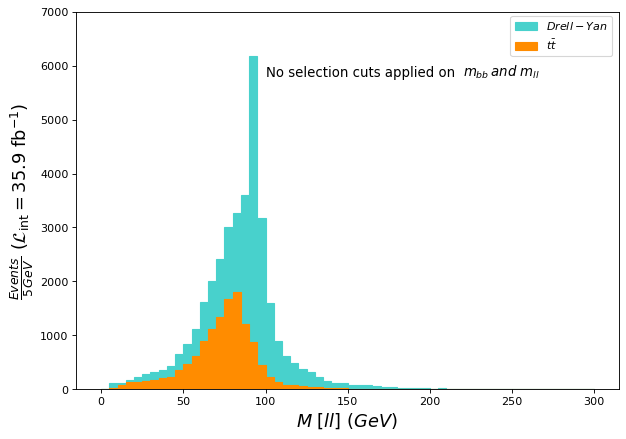}
    \includegraphics[width=0.8\linewidth]{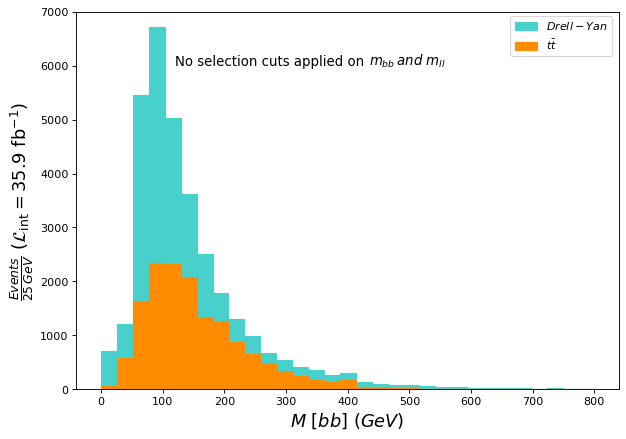}
    \caption{The background processes as a function of $m_{\ell^+\ell^-}$ and $m_{b\bar b}$ without any cuts applied on $m_{\ell^+\ell^-}$ or $m_{b\bar b}$.}
    \label{bgnocuts}
\end{figure}

\begin{figure}
    \centering
    \includegraphics[width=0.8\linewidth]{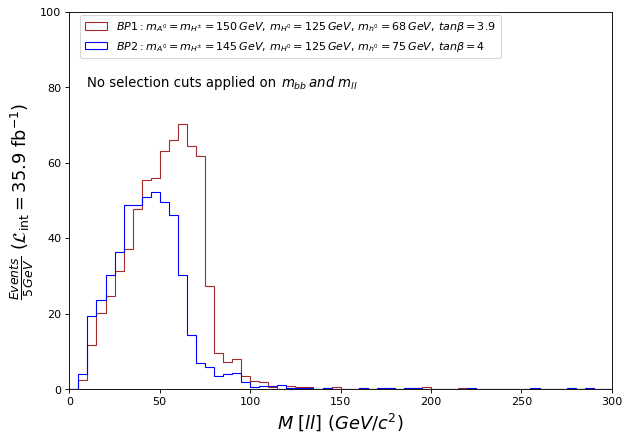}
    \includegraphics[width=0.8\linewidth]{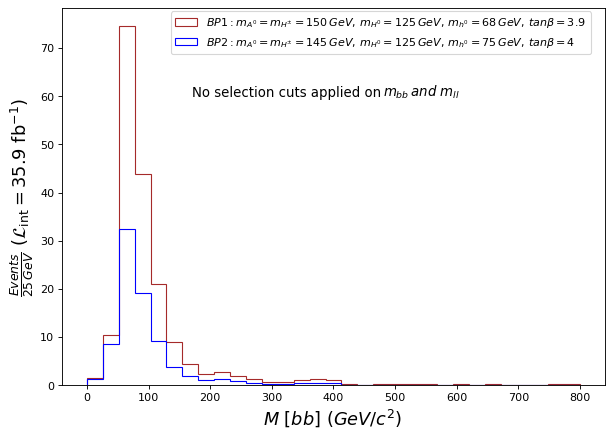}
    \caption{Signal rates as a function of $m_{\ell^+\ell^-}$ and $m_{b\bar b}$ without any cuts applied on $m_{\ell^+\ell^-}$ or $m_{b\bar b}$.}
    \label{snocuts}
\end{figure}

In order to probe the case of off-shell $Z^*$ (for which no limits are set in \cite{CMS:2019ogx} or elsewhere), it was necessary to change the  $70$ GeV $< m_{\ell^+\ell^-}< 110$ GeV cut such that more of the signal events are retained. Such a modification also has 
favorable consequences for the magnitude of the background, as we will see below.
In order to find an optimum $m_{\ell^+\ell^-}$ cut we analyzed the mass distributions $m_{\ell^+\ell^-}$ and $m_{b\bar b}$ for the DY and $t\bar{t}$ backgrounds and compared these with those of the signal. 
At this stage, no cuts were applied on $m_{\ell^+\ell^-}$ or $m_{b\bar b}$ but all other selection cuts were kept (i.e.,  $p_{T}$,  $\eta $ etc.).  For the backgrounds, in Fig.~\ref{bgnocuts}, one can see that the peak of the DY process is at $m_{\ell^+\ell^-}=m_Z$, as expected
(as the process involves a decaying on-shell $Z$), while the peak of the  $t\bar{t}$ background (in which the leptons originate from $W^\pm$ bosons) is at a lower value than $m_Z$. For the $m_{b\bar b}$ distribution in the backgrounds, the peak is at around 100 GeV
for both processes, although the $b\bar{b}$ pairs have a different origin in DY and $t\bar{t}$, being from Initial State Radiation (ISR) mistagged gluons (primarily) in the former case and from actual $b$-jets from $t\bar{t}$ decays in the latter case.
Fig.~\ref{snocuts} for the signal  illustrates that the peak in $m_{\ell^+\ell^-}$ is indeed not at $m_Z$: BP1 (which has a less off-shell $Z^*$ due to the larger value of $m_{A}-m_{h}$) having its peak at a larger value of $m_{\ell^+\ell^-}$ than BP2.
In the distribution in $m_{b\bar b}$ ones sees that the peak corresponds to the mass of $m_{H}$ (as expected, because $H\to b\bar{b}$).

Given the differences in the $m_{\ell^+\ell^-}$ and $m_{b\bar b}$ distributions for the signal and backgrounds, in order to enhance the significance of the former over the latter, we choose the cuts $20$ GeV $< m_{\ell^+\ell^-} < 50$ GeV and  $m_{b\bar b} < 100$ GeV, with the $m_{\ell^+\ell^-}$ cut causing the greater suppression of the background. These requirements were chosen based on the expected kinematic features of the signal, specifically, the presence of an 
off-shell $Z^*$ boson and a light Higgs boson (i.e., $h$ with a mass below 125 GeV). These particular values were selected to significantly reduce the dominant background contributions while retaining a significant portion of the signal. However, no formal optimization (such as a parameter scan or significance maximization) was carried out. As such, the cuts used in this study are illustrative rather than final.

\begin{figure}
\centering
  \includegraphics[width=0.8\textwidth]{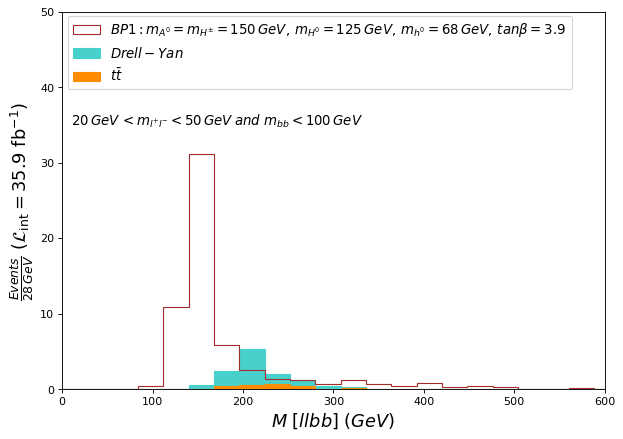}
\hfill
    \includegraphics[width=0.8\textwidth]{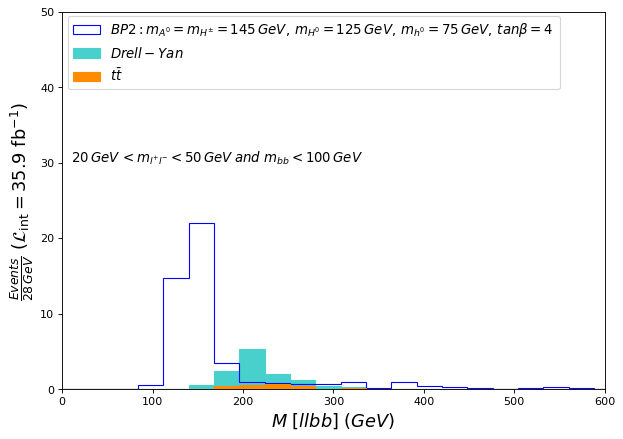}
\caption{Signal and  background processes as a function of $m_{\ell^+\ell^-b\bar b}$ for $ 20 \, {\rm GeV} \, < m_{\ell^+\ell^-} < 50\, {\rm GeV}$ and $ m_{b\bar b} < 100\, {\rm GeV}$.}
\label{bgssllbb20}
\end{figure}

In Fig.~\ref{bgssllbb20} we display the signals and backgrounds as a function of the four-body invariant mass distribution $m_{\ell^+\ell^- b\bar b}$ with the above cuts on $m_{\ell^+\ell^-}$ and $m_{b\bar b}$ being applied. One can see that the signals are now very prominent above a much smaller background,
thus allowing the case of the $A\to HZ^*$ signal to be probed. We remind the reader here that the case with $Z^*$ currently has no limits set on the $(m_A, m_{H})$ plane from the CMS search in \cite{CMS:2019ogx}.

In the remainder of our analysis, we evaluate the statistical significance  of the signals over the backgrounds using two common definitions of it. The first definition is \( \boldsymbol{\mathcal{S}}_1 = S/\sqrt{B} \), where \( S \) is the number of signal events and \( B \) is the number of background events. The second definition is \( \boldsymbol{\mathcal{S}}_2=S/\sqrt{S + B} \), which accounts for statistical fluctuations in both signal and background. Both results are reported in Tab.~\ref{s}, for the  luminosity $35.9$ fb$^{-1}$. The latter makes two key points. On the one hand, if the CMS search of \cite{CMS:2019ogx} were applied as is to our proposed signal, it would have no sensitivity to it. On the other hand, if modified suitably as recommended here, such a search would afford one with dramatic discovery possibilities. In all generality, based on previous work of ours on the size of the signal $pp\to A\to Z^*h\to \ell^+\ell^- b\bar b$ in the 2HDM Type-I in IH
\cite{Akeroyd:2023kek,Akeroyd:2024dzq,Akeroyd:2024tbp}, we conclude that such a  level of sensitivity would exist over a sizable region of its parameters space.


\begin{table}[!t]
\centering
\begin{tabular}{|p{4.8cm}|p{1.2cm}|p{1.2cm}|p{1.2cm}|p{1.2cm}|p{1.2cm}|p{1.2cm}|p{1.2cm}|}
\hline
Scenario  & $S$ \newline BP1 & $S$ \newline BP2& $B$ & ${\boldsymbol{\mathcal{S}}}_{1}$\newline BP1& ${\boldsymbol{\mathcal{S}}}_{1}$\newline BP2&
${\boldsymbol{\mathcal{S}}}_{2}$\newline BP1&
${\boldsymbol{\mathcal{S}}}_{2}$\newline BP2 \\
\hline
\small $ 70 \, {\rm GeV} \, < m_{\ell^+\ell^-} < 110 \, {\rm GeV}$ & 178 & 116 & 13924 & 1.5& 0.9&1.4 &0.86\\
\hline
\small $ 20 \, {\rm GeV} \, < m_{\ell^+\ell^-} < 50 \, {\rm GeV}$ \newline {\rm and} 
$m_{b\bar b} < 100 \, {\rm GeV}$ & 153 & 100 & 9 & 43&28& 12&9.5 \\
\hline
\end{tabular}
\caption{Signal ($S$), background ($B$) and significances (${\cal S}_1$ and ${\cal S}_2$) for two BPs where $20\, {\rm GeV} <m_{\ell^+\ell^-b\bar b} <180 \, {\rm GeV}$ with an integrated luminosity of $35.9$ fb$^{-1}$.}
\label{s}
\end{table}

\section{Conclusions}
\label{74}
In this paper, we have demonstrated that optimized selection criteria can significantly enhance the LHC observability of the signal process $pp\to A \to Z^{*}h \to \ell^+\ell^- b\bar{b}$ within the framework of the 2HDM Type-I, specifically, under the assumption of an IH wherein $m_{h}<m_H=125$ GeV. In particular, by investigating mass selection cuts alternative to the CMS baseline used in Higgs searches in the $\ell^+\ell^-b\bar b$ final state 
(wherein an on-shell $Z$ boson is sought),  
specifically, targeting the low $m_{\ell^+\ell^-}$ region and imposing an upper bound on $m_{b\bar b}$, we have shown that dominant backgrounds from the DY and $t\bar{t}$ processes can be substantially reduced. This ultimately leads to a notable improvement in the signal significance of the target signal process, whichever its  definition. Our results highlight the effectiveness of the refined kinematic selections and we thus encourage the LHC collaborations to carry out a search for $pp\to A \to Z^*h \to \ell^+\ell^- b\bar{b}$ along the lines presented here, with the aim of increasing sensitivity to the regions of the ($m_{A}, m_{h}$) plane where $m_{A} - m_{h} < m_Z$, i.e., the $Z^*$ is off shell, a kinematic configuration  not presently considered experimentally.\\

\noindent
{\bf Acknowledgements}
\vspace*{0.5cm}

SA acknowledges the use of the IRIDIS High Performance Computing Facility and associated support services at the University of Southampton. SA further acknowledges
support from a scholarship of the Imam Mohammad Ibn Saud Islamic University.
AA and SM are funded in part by the STFC CG ST/X000583/1 and 
SM also through the NExT Institute.


\end{document}